\documentclass[10pt]{article}
\usepackage{epsfig,dsfont}
\usepackage{geometry}
\begin{document}
\sffamily

\thispagestyle{empty}
\vspace*{15mm}

\begin{center}

{\LARGE 
A study of the sign problem for
lattice QCD with chemical potential}
\vskip20mm
Julia Danzer$^a$, Christof Gattringer$^a$, Ludovit Liptak$^b$, 
Marina Marinkovic$^c$
\vskip18mm
$^a\,$Institut f\"ur Physik, Unversit\"at Graz, \\
Universit\"atsplatz 5, 8010 Graz, Austria 
\vskip5mm
$^b\,$Institute of Physics, Slovak Academy of Sciences, \\
Dubravska cesta 9, 845 11 Bratislava, Slovak Republic 
\vskip5mm
$^c\,$Institut f\"ur Physik, Humboldt-Universit\"at zu Berlin  \\
Newtonstra{\ss}e 15, D-12489 Berlin, Germany
\end{center}
\vskip30mm

\begin{abstract}
We study the expectation value of the phase of the 
fermion determinant for Wilson lattice fermions with
chemical potential. We use quenched SU(3) ensembles and implement a recently 
proposed exact dimensional reduction of the fermion determinant. Ensembles
at several temperatures below and above the phase transition 
are studied and we analyze the role of the quark 
mass, the temperature, the volume and the topological sectors. 
We compare our numerical
results to predictions from chiral perturbation theory. 
\end{abstract}

\vskip20mm
\begin{center}
{\sl Final version to appear in Physics Letters B.}
\end{center}

\setcounter{page}0
\newpage
\noindent
{\Large 1. Introductory remarks}
\vskip4mm
\noindent
With the increasing amount of experimental data on the QCD phase diagram, 
corresponding ab-initio lattice calculations become more and more important. 
However, when a chemical potential is introduced, lattice simulations face a
serious challenge, the fermion sign problem. With non-zero chemical potential 
$\mu$ the fermion determinant $\det[D(\mu)]$ 
is complex and cannot be directly used as a 
probability weight. Unless conceptually new ideas are developed,
Monte Carlo simulations need to use various kinds of reweighting techniques.

The severeness of the sign problem, and thus the numerical effort for a
reweighting strategy, may be characterized by the expectation value
\begin{equation}
\Big\langle \, e^{i 2 \theta} \, \Big\rangle \; = \; \left\langle \,
\frac{\det[D(\mu)]}{\det[D(-\mu)]} \, \right\rangle \; ,
\label{phasesquare}
\end{equation}
where $e^{i\theta}$ is the phase of the fermion determinant $\det[D(\mu)]$. 
Having the extra factor of 2 in the exponent on the lhs.\ is convenient, 
since the expectation value of that phase may be written as a ratio 
of two determinants.

Recently the determinant phase (\ref{phasesquare}) was addressed in several
papers  \cite{chpt1} -- \cite{rmt2} using different analytical tools, such as 
chiral perturbation theory or random matrix models. The dependence of the sign
expectation value $\langle e^{i 2 \theta} \rangle$ on the chemical potential,
the volume, the temperature, the quark/pion mass and the topological sector
was studied. On the lattice several individual results may be spotted
\cite{lattice1} -- \cite{lattice2}, but a
systematical analysis of the sign problem is still missing. 

In this paper we attempt a small step towards a more complete analysis and
study the determinant phase for quenched ensembles in a wide range of 
temperatures $T$ and chemical potential values $\mu$. Even in the quenched 
situation this is still a sizable task, but 
applying the recently proposed factorization formula 
\cite{DaGa,canonical} we are able to 
speed up the evaluation of the determinant phase considerably. We study the
dependence of $\langle e^{i 2 \theta} \rangle$ on the parameters
$\mu$, $T$, the quark
mass $m$, the volume and the topological sectors. In particular we also
compare the behavior in the low- and high temperature phases of quenched QCD.   

\vskip7mm
\noindent
{\Large 2. Technicalities}
\vskip4mm
\noindent 
In our study we use quenched ensembles generated with the L\"uscher-Weisz
action \cite{luweact} on lattices of size $N^3 \times N_T = 6^3 \times 4$, 
$8^3 \times 4$ and $10^3 \times 4$. 
The scale was determined in \cite{scale} based on the 
Sommer parameter. We generated ensembles for a wide range of temperatures
between $T = 210$ MeV and  $T = 430$ MeV, with the critical temperature 
for our action determined \cite{tcrit} to be $T_c = 300(3)$ MeV. Our
statistics is between 500 and 2000 configurations for the smaller lattices and 
100 configurations for the larger ones. An overview of our ensembles is 
given in Table \ref{parametertable}.

\begin{table}[Ht]
\begin{center}
\begin{tabular}[t]{ccccccc}
\hline
\hline
$\;\; N^3\times N_T$ \quad  & \quad $\!\!\!\beta_G$ \quad &  
$\;$ $a$ [fm]  \quad 
&  $\;T$ [MeV] $\;$  & $\;$ $T/T_c$ $\;$ &  \quad $m$ [MeV]  
$\;$ & \# confs.\\
\hline
$6^3 \times 4$ & 7.5 & 0.213 & 232 & 0.77 & 100 & 500 \\
\hline
$8^3 \times 4$ & 7.4 & 0.234 & 210 & 0.70 & 50, 100, 200 & 2000, 2000, 1000 \\
$8^3 \times 4$ & 7.5 & 0.213 & 232 & 0.77 & 100 & 500 \\
$8^3 \times 4$ & 7.6 & 0.194 & 254 & 0.85 & 100 & 500 \\
$8^3 \times 4$ & 7.7 & 0.177 & 279 & 0.93 & 100 & 500 \\
$8^3 \times 4$ & 7.8 & 0.161 & 306 & 1.02 & 100 & 500 \\
$8^3 \times 4$ & 7.9 & 0.148 & 334 & 1.11 & 100 & 500 \\
$8^3 \times 4$ & 8.0 & 0.135 & 364 & 1.21 & 100 & 500 \\
$8^3 \times 4$ & 8.1 & 0.125 & 396 & 1.32 & 100 & 500 \\
$8^3 \times 4$ & 8.2 & 0.115 & 430 & 1.43 & 50, 100, 200 & 500, 1000, 500 \\
\hline
$10^3 \times 4$ & 7.5 & 0.213 & 232 & 0.77 & 100 & 100 \\
$10^3 \times 4$ & 8.1 & 0.125 & 396 & 1.32 & 100 & 100 \\
\hline
\hline
\end{tabular}
\end{center}
\caption{Table with the parameters of our ensembles. We list the size of the
  lattice in lattice units, 
the gauge coupling $\beta_G$, the lattices spacing $a$ in fm, the
  temperature $T$ in MeV and as a multiple of $T_c$, the bare quark mass $m$ 
in MeV
  and the statistics.}
\label{parametertable}
\end{table}

For our gauge ensembles we determined the fermion determinant for the Wilson
Dirac operator $D(\mu)$ with chemical potential $\mu$ (the lattice spacing is
set to $a = 1$ here),
\begin{equation}
D(\mu)_{x,y} \; = \; \delta_{x,y}  \; - \; \kappa 
\sum_{\nu = \pm 1}^{\pm 4} 
\; e^{\pm \mu \, \delta_{|\nu|,4}} \; 
\frac{ \mathds{1} \mp \gamma_{|\nu|}}{2} \; U_\nu(x) \; \delta_{x +
  \hat{\nu},y} \; . 
\label{wilsondirac}
\end {equation}
We use the convention $U_{-\nu}(x) = U_\nu(x-\hat{\nu})^\dagger$, introduce the
abbreviation $\hat{\nu}$ for the shift vector in the $\nu$ direction, and 
$\kappa$ is related to the bare mass $m$ via $\kappa = 1/(4+m)$. The temporal
boundary conditions for the fermions are anti-periodic, all other boundary
conditions (gluons, spatial boundary conditions for the fermions) are
periodic. 

In order to speed up the evaluation of the fermion determinant for many values
of $\mu$ at the same time, we use the dimensional reduction formula developed
in \cite{DaGa}. We here only very briefly sketch the idea of the
construction and refer to \cite{DaGa,canonical} for the technical details:
Applying a decomposition of the lattice into four domains, 
the fermion
determinant may be rewritten in the form 
\begin{equation}
\det[D(\mu)] \; = \; A_0 \, \det\! \Big[ \mathds{1} \, - \, H_0 \, - \,
e^{\,\mu N_T} H_{+1} \, - \, e^{- \mu N_T} H_{-1} \Big] \; .
\label{determinantfactor}
\end{equation}
Here $A_0$ is a factor that is essentially a product of determinants 
for the terms of the Dirac operator restricted to the four domains. 
This factor is real and independent of the chemical potential $\mu$. 
All of the $\mu$-dependence comes from the second factor which has again the
form of a determinant. However, the matrices $H_0, H_{\pm 1}$ live on 
only a single time slice and thus the evaluation of this second determinant 
is by a factor of $N_T^3 = 4^3 = 64$ cheaper than the evaluation of the
determinant in the original formulation. The matrices $H_0, H_{\pm 1}$
are made from products of propagators on the domains and are small 
enough such that they can be stored in memory. Then with
(\ref{determinantfactor}) the determinant $\det[D(\mu)]$
can be evaluated for several values of $\mu$ very efficiently. Actually, since 
$A_0$ cancels in (\ref{phasesquare}), 
for the phase factor $e^{i2\theta}$ only the dimensionally 
reduced determinant $\det [ \mathds{1} -  H_0  - 
e^{\,\mu N_T} H_{+1}  -  e^{- \mu N_T} H_{-1} ]$ is needed. We typically 
use 16 to 26 different values for $\mu$ spaced with $\Delta \mu = 0.05$ (in
lattice units). 

For some of our ensembles we also evaluated the topological charge of the
configurations. For that purpose the low lying eigenvalues of the 
overlap operator were computed and the topological charge was determined from
the number of zero modes using the index theorem.

\begin{figure}[t]
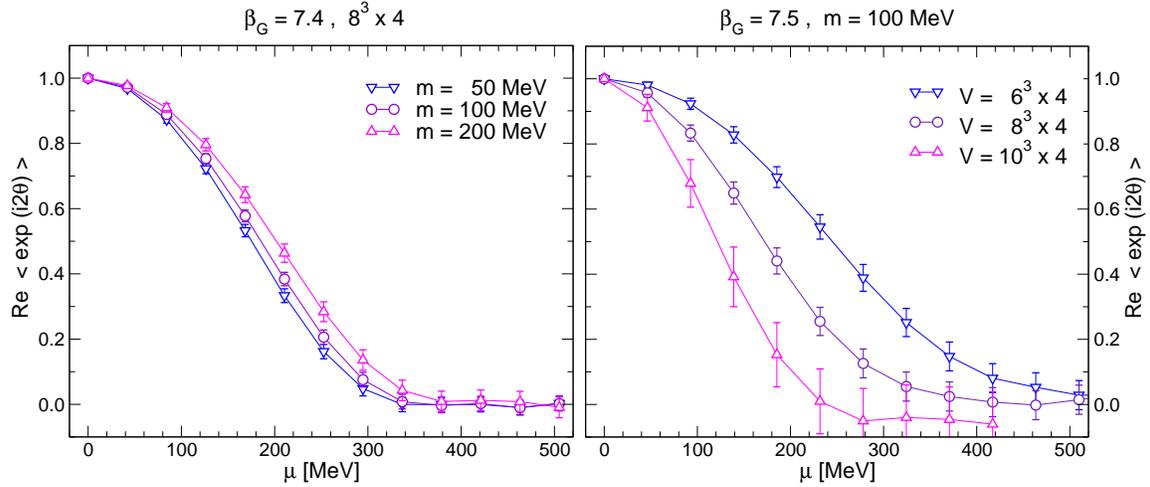

\begin{center}
\includegraphics[width=75mm,clip]{Re_exp2theta_masses_8x4_b740.eps}
\includegraphics[width=75mm,clip]{Re_exp2theta_vsV_b750.eps} 
\end{center}
\caption{Phase of the determinant as a function of the chemical potential for
  low temperature. In the lhs.\ plot we compare the dependence on the quark
  mass in a fixed volume, while in the rhs.\ plot the mass is held fixed and
  we vary the lattice size. The symbols are connected to guide the eye.} 
\label{phase_vs_mu} 
\end{figure}

\vskip7mm
\noindent
{\Large 3. Numerical results}
\vskip4mm
\noindent 
{\large 3.1 Qualitative dependence on mass and volume}
\vskip4mm
\noindent 
We begin our presentation of the numerical results with a qualitative
discussion of the behavior of $\langle e^{i 2 \theta} \rangle$ as a function
of the quark mass and the volume. In the lhs.\ plot of Fig.~\ref{phase_vs_mu}
we show $\langle e^{i 2 \theta} \rangle$ for the $8^3 \times 4$ ensemble at 
gauge coupling $\beta_G = 7.4$ where we have three values of the bare quark
mass, $m = 50, 100$ and 200 MeV. For all three values we observe a
Gaussian-type of distribution. The width of the distribution depends on the
mass with the largest mass giving the widest distribution, in other words 
small mass increases the sign problem.  

Similarly in the rhs.\ plot of Fig.~\ref{phase_vs_mu} we compare the three
volumes which we have at $\beta_G = 7.5$ for $m = 100$ MeV. 
Here it is obvious, that the 
distribution is wider for small volume, i.e., increasing the volume makes the
sign problem more severe. 

We remark that although we here only show plots for 
ensembles in the low temperature phase, we confirmed that the same behavior is
also found in the high temperature phase, i.e., small mass and large volume 
increase the sign problem. This subsection is only meant to qualitatively
demonstrate the behavior of $\langle e^{i 2 \theta} \rangle$ as a function
of the quark mass and the volume. However, at least in the low temperature 
phase one can go beyond that and compare the numerical results to quantitative
analytical predictions. This will be done in the next section. 

\begin{figure}[t]
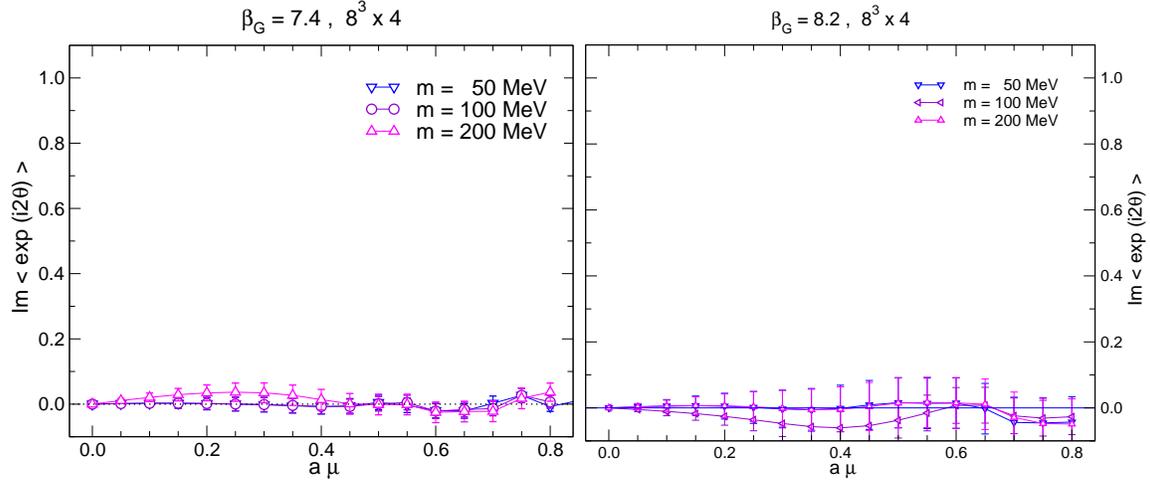

\begin{center}
\includegraphics[width=75mm,clip]{Im_exp2theta_masses_8x4_b740.eps}
\includegraphics[width=75mm,clip]{Im_exp2theta_masses_8x4_b820.eps}
\end{center}
\caption{Imaginary part of the determinant phase as a function of the chemical
  potential. We show results for a low- and a high temperature ensemble on
  $8^3 \times 4$ at three different quark masses. The symbols are connected to
  guide the eye.}
\label{Imphase_vs_mu} 
\end{figure}

In this section we conclude with a check of the reliability of our
numerical results. Since the phase of the determinant is a difficult to
measure quantity -- after all it is exponentially suppressed for large $\mu$
-- such a consistency check is important. In principle 
the phase $e^{i 2 \theta}$ is a complex number. However, due to
the symmetry under time reflections $\langle e^{i 2 \theta} \rangle$ is real. 
Thus a check for the quality of the Monte Carlo result for 
$\langle e^{i 2 \theta} \rangle$ is to inspect the imaginary part of that
quantity and to control whether it is compatible with zero. In 
Fig.~\ref{Imphase_vs_mu} we show the imaginary part of the phase expectation
value for two ensembles in the low- and the high temperature phases for three
different quark masses. In all cases we find that the imaginary part is zero 
within error bars as expected.

\vskip7mm
\noindent 
{\large 3.2 Comparison to chiral perturbation theory}
\vskip4mm
\noindent 
After the first round of a more qualitative assessment of $\langle e^{i 2
  \theta} \rangle$ in the last subsection, we now focus on the low temperature
regime where analytical results are available. Since for our ensembles we have 
$m_\pi L$ larger than 5 for all our ensembles\footnote{The lattice extent 
$L$ is given by $L = a N$, with $a$ and $N$ listed in 
Table~\ref{parametertable}. The pion masses $m_\pi$ were obtained from the
fits to chiral perturbation theory as discussed in this subsection.},
we can compare to results obtained in the $p$-regime of chiral perturbation
theory \cite{chpt1,chpt2}, while we cannot expect agreement with the 
random matrix theory results in the microscopic regime.  

The basis for our comparison are the chiral perturbation theory results for
$\langle e^{i 2  \theta} \rangle$ discussed in \cite{chpt2}. In particular the
result for the expectation value of the determinant phase in the quenched case
reads
\begin{equation}
\langle \, e^{i 2 \theta} \, \rangle  \; = \; e^{\,-\,g_0(\mu) \, + \,
  g_0(0)} \; \; \; , \; \; \; 
g_0(\mu) \; = \;  \frac{V \, m_\pi^2}{\pi^2 \, \beta^2}\; \sum_{n=1}^\infty
\frac{1}{n^2} \, K_2(n \, m_\pi \beta) \, \cosh(n \, 2\mu \beta) \; .
\label{chptresult}
\end{equation}
Here $V$ is the 4-volume, $\beta = 1/aN_T$ is the inverse temperature, 
and $K_2$ denotes the modified Bessel function. 
The expression (\ref{chptresult}) is expected to describe the phase
expectation value for $\mu < m_\pi/2$. 

\begin{figure}[t]
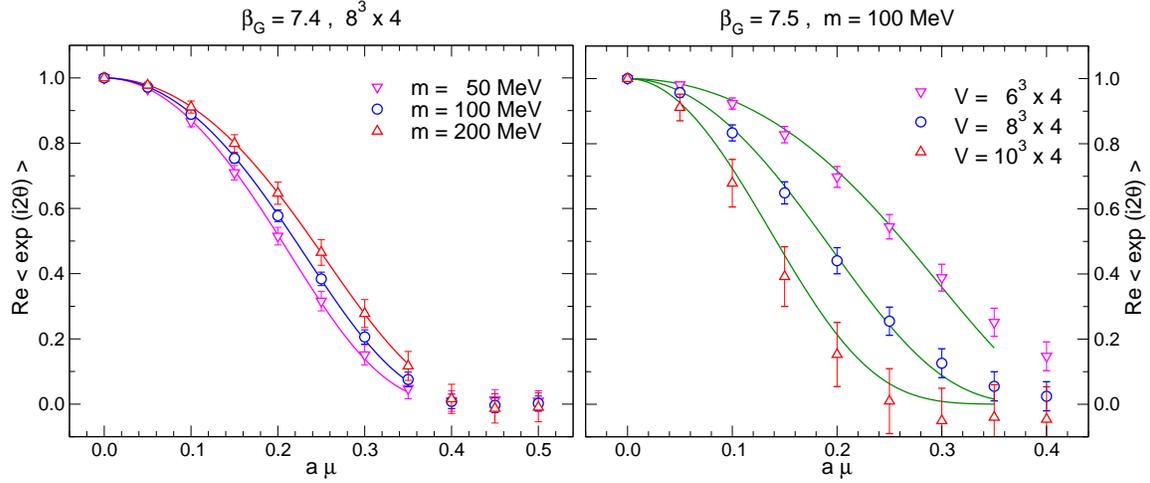

\begin{center}
\includegraphics[width=75mm,clip]{fit_8x4_masses.eps}
\includegraphics[width=75mm,clip]{fit_vol.eps} 
\end{center}
\caption{Phase of the determinant plotted as a function of $\mu$ in
  lattice units for
  low temperature. In the lhs.\ plot we compare the dependence on the quark
  mass in a fixed volume, while in the rhs.\ plot the mass is held fixed and
  we vary the lattice size. The symbols are our numerical data, and the full
  curves represent the fits to the chiral perturbation theory 
results.}
\label{phase_fits} 
\end{figure}

In Fig.~\ref{phase_fits} we show again the results for the low temperature 
ensembles already used in Fig.~\ref{phase_vs_mu}, but now plot them as a 
function of $\mu$ in lattice units (which is more convenient for our fits). 
In the lhs.~plot we performed individual one parameter (the pion mass 
$m_\pi$) fits for all three quark masses. 
The fit parameters for the three ensembles with bare quark
masses  $m = 50, 100, 200$ MeV, are $a m_\pi = 0.812(25), 0.861(25)$ and 
$0.938(34)$, which corresponds to physical pion masses of 
$m_\pi = 682(21), 724(21)$ and
$789(29)$ MeV \footnote{We remark at this point, that the quark masses $m$ 
we quote
  are bare quark masses for a non-chiral lattice Dirac operator (Wilson
  operator), and one cannot expect $m_\pi^2 \propto m$ here.}. Only data points with $\mu < m_\pi/2$, where (\ref{chptresult})
is expected to hold, were taken into account in the fit. 
It is obvious that the corresponding curves in the lhs.\ plot of 
Fig.~\ref{phase_fits} describe the data very well.
Actually the values of $\chi^2 / d.o.f.$ are rather small (below 0.1)
throughout.

The data in the rhs.\ plot of Fig.~\ref{phase_fits} allow for a much more
stringent test of the chiral perturbation theory result (\ref{chptresult}). The 
three ensembles used there differ only by their volume, while the gauge 
coupling $\beta_G$ and the quark mass $m$ were held fixed. Thus 
one expects that the three ensembles have the same pion mass 
(neglecting possible finite size effects). Since the volumes $V$ are known
for the three ensembles, one can attempt a single one parameter 
(the pion mass $m_\pi$) fit for all three volumes simultaneously. The result
of this fit is shown by the full curves in the rhs.\ plot of 
Fig.~\ref{phase_fits} and obviously represents the data pretty well. 
The outcome for the fit parameter 
is $a m_\pi = 0.729(52)$ which corresponds to a pion mass of $m_\pi = 675(49)$ 
MeV (note that here the gauge coupling is different from the one in the 
lhs.\ plot). We find $\chi^2 / d.o.f. = 0.485$ which demonstrates that 
 (\ref{chptresult}) describes the data for the different volumes 
very well. It is also interesting to note that the result for the pion mass
from the fit ($m_\pi = 675$ MeV) 
is rather close to what one would estimate from the quenched 
spectroscopy calculation \cite{bgr} done for the L\"uscher-Weisz action   
($m_\pi = 660$ MeV). 

\vskip7mm
\noindent 
{\large 3.3 The role of the topological charge }
\vskip4mm
\noindent 
An interesting effect has been discussed in \cite{rmt2}: In a random matrix
calculation in the microscopic regime it was found that the phase 
expectation value $\langle e^{i 2  \theta} \rangle$ should depend on the
topological charge $Q$ of the gauge configurations. The distribution of
 $\langle e^{i 2  \theta} \rangle$ becomes wider as $|Q|$ is increased, in
other words, the sign problem is milder for higher charge sectors. The
question which we address here is whether the topological effect 
is a specific feature of the microscopic regime or plays a role in general.

\begin{figure}[t]
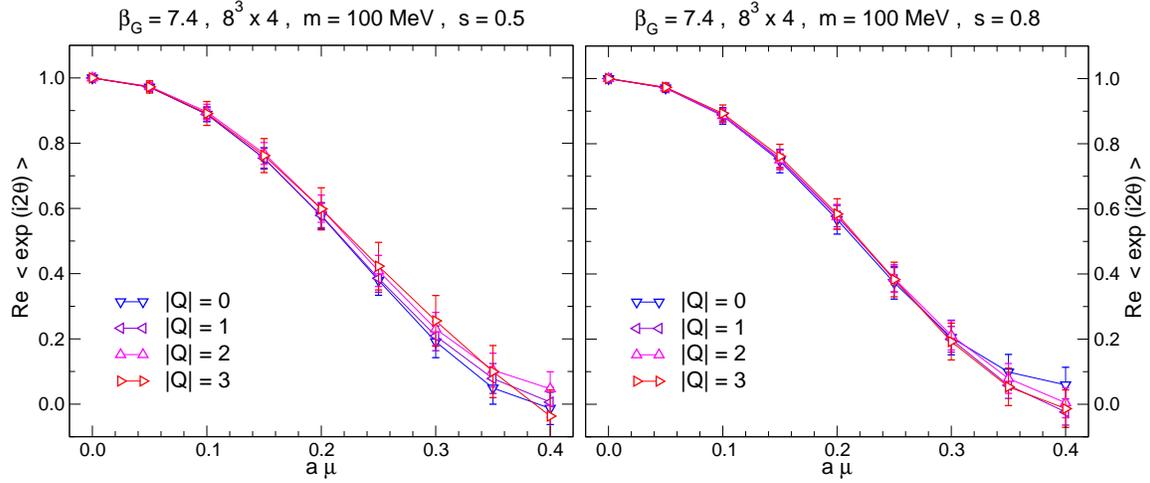

\begin{center}
\includegraphics[width=75mm,clip]{Re_exp2theta_topch05_8x4_b740.eps}
\includegraphics[width=75mm,clip]{Re_exp2theta_topch08_8x4_b740.eps}
\end{center}
\caption{Phase of the determinant as a function of $\mu$ in lattice units for
our $\beta_G = 7.4, 8^3 \times 4$ ensemble. In this plot we separate the data
with respect to the topological charge $Q$ of the gauge configurations. We 
compare $s = 0.5$ (lhs.\ plot) and $s = 0.8$ (rhs.).}
\label{phase_vs_mu_topcharge} 
\end{figure}

As already remarked, we determine the topological charge via the number of
zero modes of the overlap Dirac operator using the index theorem. The overlap
operator has a free parameter $s$ which may be used to tune the locality
\cite{locality}. This parameter, however, also influences the number
of zero modes as it shifts the center of the overlap projection. We
compare two different values $s = 0.5$ and $s = 0.8$ which give slightly
different results for the topological charge. The agreement of the topological
charges becomes better as $\beta_G$ is increased. 

In Fig.~\ref{phase_vs_mu_topcharge} we show $\langle e^{i 2  \theta} \rangle$ 
as a function of the chemical potential in lattice units for
our $\beta_G = 7.4, 8^3 \times 4$ ensemble, separating the
gauge ensembles with respect to the topological charge $Q$. The lhs.\ plot is
for $s = 0.5$, the rhs.\ for $s = 0.8$. The plots clearly indicate that there 
is no topological effect for $\langle e^{i 2  \theta}
\rangle$ in our data. 
The curves are on top of each other within error bars, and the
central values do not show the expected monotony (upwards trend with
increasing $|Q|$). The same absence of the topological effect was also found
for the other values of $\beta_G$. One may speculate about the reason for the
absence of the topological effect. It could well 
be that it is indeed seen only in
the microscopic regime or only for a chirally symmetric Dirac operator 
such as the overlap, but of course also the analysis could be improved by
using larger and finer lattices, where the topological charge is better 
defined. Since the cost for 
evaluating $\langle e^{i 2  \theta} \rangle$ rises tremendously with the 
volume, such a study must be left for the future.

\begin{figure}[t]
\begin{center}
\includegraphics[width=135mm,clip]{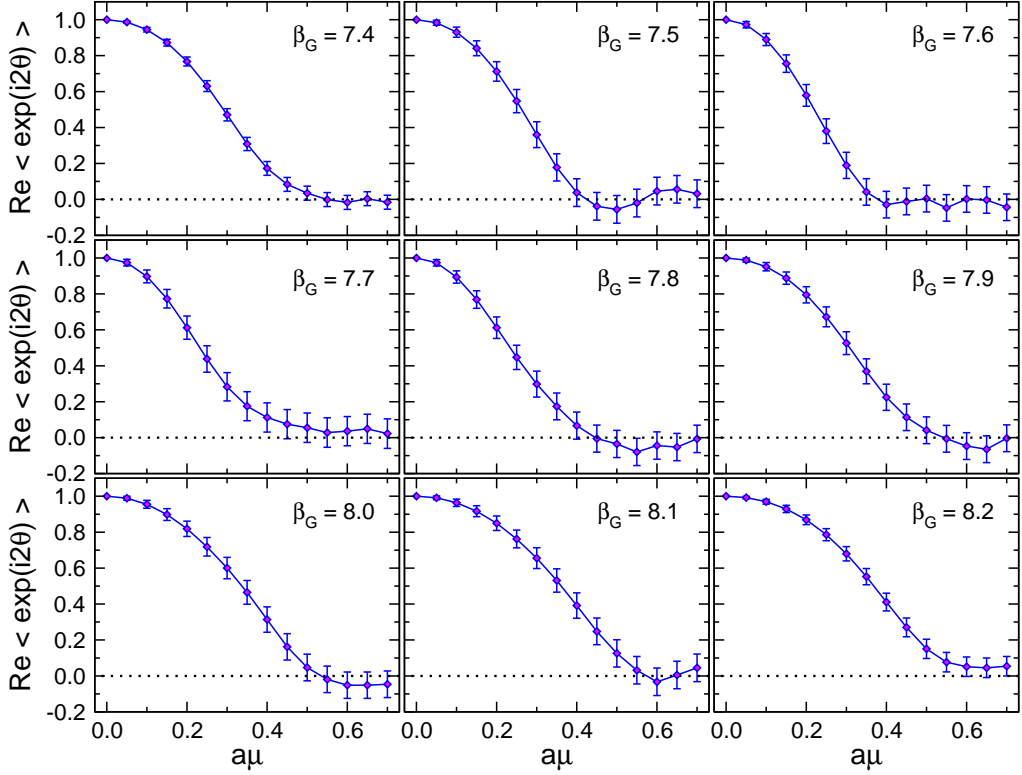} 
\end{center}
\caption{Phase of the determinant 
as a function of the chemical potential in lattice units. We present the
results for all values of the gauge coupling on our $8^3\times4$ ensembles. 
For these plots we only use the configurations in the real center
  sector.} 
\label{phase_allT} 
\end{figure}

\vskip7mm
\noindent 
{\large 3.4 Increasing the temperature}
\vskip4mm
\noindent      
We now discuss the behavior of $\langle e^{i 2  \theta} \rangle$ as one 
increases the temperature. For the quenched theory the system undergoes a phase 
transition at $T_c \sim 300$ MeV, where the center symmetry is broken 
spontaneously. This breaking is, e.g., signaled by the Polyakov loop which 
vanishes below $T_c$ and has a finite value above $T_c$. In the high
temperature phase the system spontaneously selects one out of three possible
center sectors, which are distinguished by the phase $\phi_P$
of the Polyakov loop, $\phi_P
\sim 0, 2\pi/3$ or $-2\pi/3$. The fermion determinant on the other hand is not
invariant under center transformations and becomes very small for the two 
complex sectors due to self averaging of the canonical determinants in the
fugacity expansion \cite{canonical}. Only for the real center sector the
fermion determinant remains large, which in turn is the reason that in a 
dynamical simulation always the real center sector is selected. 
Consequently in our analysis we here consider the real center sector. 
We only take into account those gauge configurations where the phase $\phi_P$ 
of the Polyakov loop obeys $|\phi_P| < \pi/3$. For the other sectors 
the fermion determinant becomes relatively small and the phase factor 
$e^{i2\theta}$ is ill defined.

\begin{figure}[t]
\begin{center}
\includegraphics[width=80mm,angle = -90,bb= 100 80 520 750,clip]{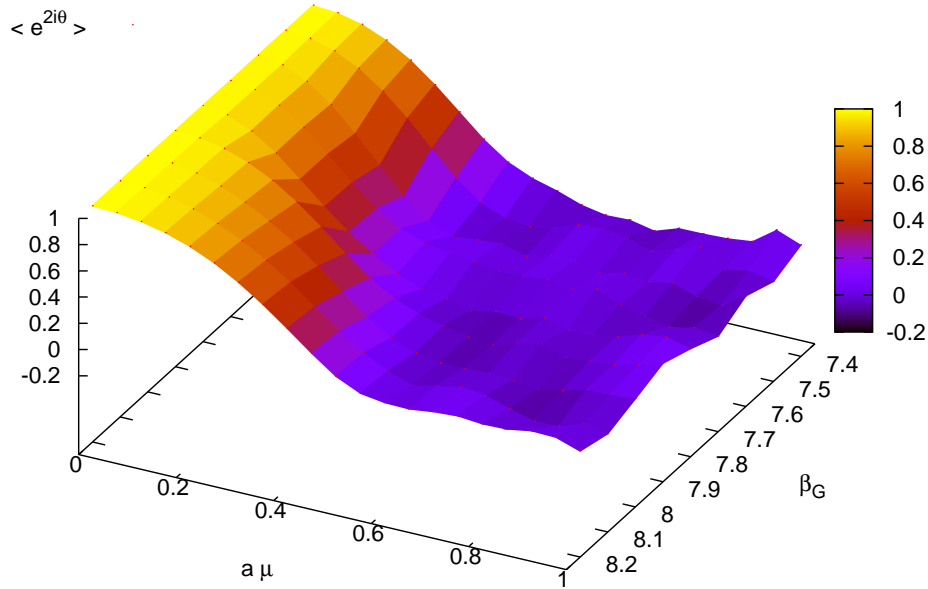} 
\end{center}
\caption{3-D plot of the phase of the determinant as a function of the
  chemical potential in lattice units and the gauge coupling. The data are for
the $8^3\times4$ ensembles restricted to the real center loop sector.} 
\label{phase3d} 
\end{figure}

In Fig.~\ref{phase_allT} we give an overview of the determinant phase for
the $8^3 \times 4$ ensembles plotted
as a function of the chemical potential in lattice units, comparing all our
values of the gauge coupling. The critical value for the transition is roughly
given by $\beta_G = 7.8$. In Fig.~\ref{phase3d} the same information is
presented as a 3-d plot. The figures show clearly that qualitatively the 
distribution of $\langle e^{i 2  \theta} \rangle$ versus $\mu$ keeps the
Gaussian type of shape as the temperature is increased. The distribution
first seems to become more narrow with increasing $\beta_G$, but 
from $\beta_G = 7.8$ on widens again. 
However, it is important to keep in mind, that although in lattice units the 
volume remains fixed, the lattice spacing and thus the physical volume
decrease with increasing $\beta_G$. One expects that the shrinking physical 
volume has a mildening effect for the sign problem (see Subsections 3.1 and
3.2). The main message of  
Figs.~\ref{phase_allT} and \ref{phase3d} thus is of more qualitative nature: 
At the quenched transition there is only a gradual change in the distribution 
of $\langle e^{i 2  \theta} \rangle$ and we do not observe a dramatic
qualitative effect for the sign problem. 

\vskip7mm
\noindent
{\Large 4. Discussion}
\vskip4mm
\noindent 
In this paper we have made a step towards a more systematical understanding
of the fermion sign problem for lattice QCD with chemical potential. Using
quenched ensembles the phase 
of the fermion determinant is analyzed for a wide range of temperatures below
and above the phase transition and for several values of the chemical
potential. We analyze the dependence of the determinant phase on the
temperature, the chemical potential, the quark mass, as well as the topological
charge and compare our results to chiral perturbation theory. 

We find that for all temperatures and values of the chemical potential the
sign problem becomes harder with increasing volume and decreasing quark mass. 
In general the determinant phase has a Gaussian type of distribution as a
function of the chemical potential for all temperatures we considered. At the 
deconfinement transition of the quenched theory we do not observe any
dramatic qualitative
effect for the sign problem. Concerning a possible dependence on the 
topological charge, we do not find such a topological effect in the regime
we work at. However, here technical improvements, in particular 
larger and finer lattices, where the concept of topological charge is better 
defined, would be needed for a final answer on the fate of the topological
effect outside the microscopic regime.  

\vskip10mm
\noindent
{\Large Acknowledgments}
\vskip4mm
\noindent
We thank Jacques Bloch, 
Christian Lang, Stefan Olejnik and Kim Splittorff 
for inspiring discussions. 
J.~D.\ acknowledges support by the FWF DK grant number W-1203, and M.~M.\ 
support from the Austrian Agency for International Cooperation in Education
and Research OeAD. Furthermore this research was supported in 
part by the Slovak Grant
Agency for Science, Project VEGA No.\ 2/0070/09, by ERDF OP R$\&$D,
Project CE QUTE ITMS NFP 262401022, and by the Center of Excellence SAS
QUTE (L.L.). The numerical computations 
were mainly performed on the cluster of the
Department of Complex Physical Systems (Institute of Physics, Bratislava), and
partly at the ZID, University of Graz.

\end{document}